\documentclass[preprint,aps]{revtex4}
\usepackage{graphicx}
\usepackage{color}
\makeatletter
\usepackage{epsfig}
\def\@dotsep{4.5}
\usepackage{dcolumn}
\usepackage{amsmath}
\makeatother

\graphicspath{{./figures/}}

\usepackage{graphicx}
\usepackage{bm}

\usepackage{amsfonts}
\usepackage{amsbsy}
\usepackage{amssymb}
\usepackage[mathscr]{eucal}

\newcommand{\beq}{\begin{equation}}
\newcommand{\eeq}{\end{equation}}
\newcommand{\ba}{\begin{array}}
\newcommand{\ea}{\end{array}}
\newcommand{\bea}{\begin{eqnarray}}
\newcommand{\eea}{\end{eqnarray}}
\newcommand{\bseq}{\begin{subequations}}
\newcommand{\eseq}{\end{subequations}}

\date{today}

\begin{document}

\title{Quantum reflection of rare gas atoms and clusters from a grating}

\author{G. Rojas-Lorenzo and J. Rubayo-Soneira}
\email{german@instec.cu,jrs@instec.cu}
\affiliation{Instituto Superior de Tecnolog\'ias y Ciencias Aplicadas (InsTEC), Universidad de La
Habana, Ave. Salvador Allende No. 1110,  Quinta de Los Molinos, La Habana 10400, Cuba \\}
\author{S. Miret-Art\'es}
\email{s.miret@iff.csic.es}
\affiliation{Instituto de F\'isica Fundamental, Consejo Superior de Investigaciones
Cient\'ificas, \\ Serrano 123, 28006 Madrid, Spain \\}
\author{E. Pollak}
\email{eli.pollak@weizmann.ac.il}
\affiliation{Department of Chemical and Biological Physics, Weizmann Institute of Science, 76100,
Rehovot, Israel}

\date{\today}

\begin{abstract}
Quantum reflection is a universal property of atoms and molecules when scattered from surfaces in ultracold collisions. Recent experimental work 
has documented the quantum reflection and diffraction of He atoms, dimers, trimers and Neon atoms when reflected from a grating. Conditions 
for the observation of emerging beam resonances have been discussed and measured. In this paper, we provide a theoretical simulation of the 
quantum reflection from a grating for those systems. We confirm the universal dependence on the incident de Broglie wavelength  
with the threshold angles where the emerging beam resonances are observed. However, the angular dependence of the reflection efficiencies, that is, 
the ratio of scattered intensity into specific diffraction channels relative to the total intensity is found to be dependent on the details of the particle 
surface interaction. 

\end{abstract}

\vspace{2cm}

\maketitle


\newpage

\section{Introduction}

Optical effects (such as, for example, refraction, diffraction and interferometry) observed in matter waves
instead of light are determined by the interaction between the corresponding particles and measuring devices.
The particle specific interaction leads to some distortion and reduction of their visibility. The so-called Talbot effect  \cite{berry} or near field
interference effect is a good illustration for the differences between optical and matter wave effects. Quantum carpets are observed in the near 
field when gratings are illuminated by coherent light and particle beams. However, the patterns observed when using particles is distorted as 
compared to the same effect when using photons due to the particle grating specific interaction (Talbot-Beeby \cite{salva} ) which 
is typified by  a well in the interaction potential close to the surface of the grating.
When heavy particles such as big molecules and clusters are used, diffraction patterns are governed  mainly
by van der Waals interactions and a strong reduction of the fringe visibility is observed. \cite{arndt}

It is thus of interest to reduce such matter wave related distortions  as much as possible.
Recently, Zhao {\it et al.} \cite{wieland1} have proposed and demonstrated  that this is possible
thanks to  the well known effect of {\it quantum reflection} which is a key effect in many cold and
ultra-cold gas-phase collisions as well as the scattering of particles by solid surfaces. Lennard-Jones and
Devonshire \cite{lennard} first recognized this behavior in the atom-surface context and Kohn \cite{kohn}
showed later on that quantum reflection leads to a zero sticking probability at threshold
pointing out that this effect is a clear quantum interference effect between the incoming and reflected waves.

It is well understood that at threshold the maximum of the scattered wavefunction is observed far away from the grating, due to the very long 
de Broglie wavelength of the particle at the low energies involved. This has led to the notion that quantum reflection takes place far away from the grating/surface, with a distance of
typically tens or hundreds of nanometers where the surface-induced forces are too weak to dissociate
fragile bonds such as in He$_2$ \cite{wieland2} and one cannot consider a classical turning point for such
reflection, which would immediately dissociate a weakly bound molecule such as the He dimer. \cite{friedrich1,pasquini} This large distance 
then presumably weakens the distortions in the diffraction patterns due to the short range interactions with the surface.
In this context, Zhao {\it et al.} \cite{wieland1}, in their experimental study of the scattering of He, He$_2$ and D$_2$ by an echelette or 
blazed ruled grating, observed a "universal" dependence of  the so-called emerging beam resonances
(or threshold resonances in atom-surface scattering \cite{salva1}) which occur when a diffraction channel
just becomes open or closed. The "universality" expresses itself in the fact that the threshold depends only on the de Broglie wavelength 
of the incident beam but not on the characteristics of the particle surface interaction. The threshold incident angles are called Rayleigh 
angles. In this same study it was claimed that not only the threshold was universal but also the incident angle dependence of the "efficiency" 
(the ratio of the diffraction peak to the total intensity) is universal.

As discussed more recently \cite{salva2,jakob}, the fact that the maximal density in the scattered wavefunction appears far away from the surface does not mean that the details of the quantum reflection really are "universal" and independent of the moiety observed.
Quantum reflection is mainly governed by the long-range attractive van der Waals (vdW)-Casimir potential
tail which falls off faster than $r^{-2}$. \cite{friedrich2}. 
Senn \cite{senn} showed that, for general one-dimensional potentials
which vanish as the coordinate goes to $\pm \infty$, the reflection probability goes to unity at threshold
conditions except when the potential supports a zero energy resonance state.
The reflection coefficient decreases from unity as the incident kinetic energy increases according to $|R| \sim 1 - 2 k b \sim \exp(- 2 k b)$, 
where $k$ is the incident wave vector and $b$ a characteristic length which depends on the specifics of the particle grating interaction.
This universal behavior  is a direct result of boundary conditions and
continuity of the wave function and its derivative \cite{senn,jakob}.

In the semiclassical framework however, the analysis of quantum reflection has concentrated on the fact that in the regime of linear dependence on $k$ there is a failure of the semiclassical description of the scattering dynamics. \cite{friedrich3}
Far away
from the grating/surface the long range attractive potential exhibits a region in which the
local de Broglie wavelength is not slowly varying invalidating a semiclassical description in this so called "badlands" region of the interaction potential. Quantum reflection was thus associated with this badlands region where "quantality" is high. \cite{barnea17} For He atoms,
the badlands region of the potential is typically located at distances of several hundreds  of
atomic units from the grating/surface and thus would depend only on the long range part of the interaction potential.

Recently,  we have shown theoretically  \cite{salva2} that the quantum  reflection of He atoms from
a grating is determined not only by the long range interaction potential but also by its short range properties.
We emphasized that this short range region  is critical for obtaining theoretical reflection probabilities
and  diffraction patterns which are in fairly good agreement with the experimental results. These calculations
were carried out by using the close-coupling (CC) formalism \cite{salva3}, which is numerically
exact when convergence is reached. To distinguish between quantum reflection and "classical" reflection due to the turning
point of the repulsive part of the interaction, absorbing boundary conditions which prevented the classical reflection from occurring 
were employed. In Ref. \cite{jakob} we also showed that the badlands region of the
interaction potential is immaterial since the wavelength of the scattering particles at
threshold is much longer than the spatial extension of the badlands region. In other words, this region
does not provide a qualitative guide to the occurrence of quantum threshold reflection.

Quantum reflection thus presents theory with a number of challenges. One is to show that measured quantum reflection probabilities may be 
simulated by theory using the well known long range interaction potential but also considering the periodicity of the grating. A second challenge 
is to display the universality of the Rayleigh angles and their dependence on the incident wavelength only. A third question is the extent 
of this universality, does it also include the dependence of the efficiency on the incident scattering angle.

The purpose of this present work is to answer these challenges. To set the stage we show that with reasonable interaction potentials it is 
possible to simulate rather well the quantum reflection of He, He$_2$, He$_3$ and Ne on a "standard" or regular grating. Comparison 
with experiment 
where possible is good, the quantum reflection of He$_2$ has only been measured for a blazed grating so that the theoretical results for He$_2$
 have yet to be validated experimentally. As expected, for all systems considered we observe the universality of the Rayleigh angles in their 
 dependence on the incident wavelength only. However, the dependence of the efficiency on the incident angle is not universal and does depend 
 on the specifics of the interaction. Although we are considering here only a "standard" grating rather than the ruled grating used in the 
 experiments \cite{wieland1} this specificity is general and a consequence of the theory of quantum reflection and its dependence 
 on energy in the linear regime. In Section II we review the theory needed to implement the computations, in Section III we present the results. 
 The paper ends with a discussion of the various aspects of the quantum reflection phenomenon.

\section{Theory}

The experiment we want to analyze  has been described in detail in Ref. \cite{zhao1}.
The reflection grating is assumed to be in the  $x$-direction and consists of a 56-mm-long
microstructured array of 110-nm-thick, 10 $\mu$m-wide and 5-mm-long parallel chromium
strips on a flat quartz substrate. The period of the strips $d$ is 20 $\mu$m. With this geometry,
the quartz surface between the strips is  completely shadowed by them for all the incidence
angles used. Quantum reflection probabilities as well as diffraction patterns were measured at
different  source temperatures $T_0$ (ranging typically from 8 K up to 40 K) and pressures
around $P_0 = 6-8$ bar. In the cryogenic free jet expansion of incident particles, the kinetic
energy is given by $E_i = (5/2) k_B T_0$ where $k_B$ is the Boltzmann constant .\cite{toennies}
The incident grazing angle $\theta_i$ is usually varied between 0.4 and 15 $mrad$ and
measured with respect to the grating surface plane. The diffraction
angles $\theta_n$ are given by the conservation of the momentum or Bragg's law
\begin{equation}\label{bragg}
cos \theta_i - cos \theta_n = \frac{n \lambda}{d}
\end{equation}
where $\lambda$ is the de Broglie wave length of the incident particle and the diffraction
order is given by $n$. Negative diffraction orders correspond to diffraction angles close
to the surface grating, that is, energy in the perpendicular direction is transferred to the parallel direction. 
Final results are very often plotted as a function of the corresponding
perpendicular wave vector along the $z$-direction
\begin{equation}\label{kperp}
k_{perp} \simeq  \frac{\sqrt{5 m k_B  T_0}}{\hbar} \, sin \theta_i
\end{equation}
where $m$ is the atomic mass of the incident particle. When considering clusters such as He$_2$ and He$_3$, this expression is rewritten as
\begin{equation}\label{kperp}
k_{perp} \simeq N \frac{\sqrt{5 m k_B  T_0}}{\hbar} \, sin \theta_i
\end{equation}
with $N=2,3$, respectively \cite{peter}.

As previously used \cite{salva2}, the two-dimensional model potential between the incoming particles
and the grating is assumed to be a product of two functions
\begin{equation}
U(x,z) = V(z)\cdot h(x)
\label{Upot}
\end{equation}
where $V(z)$ describes the interaction along the perpendicular coordinate $z$ and $h(x)$
the periodic grating along the horizontal coordinate $x$.  In all our computations we do not take into consideration 
internal motion of the approaching molecule which is considered to be structureless. The first factor is taken to be
a Morse potential, $V_M(z)$, at short distances, and an attractive van der Waals-Casimir
tail $V_C$, at large distances
\begin{eqnarray}
 V(z) & = &
  \left\{ \begin{array}{ll}
   V_M(z)=D \left[  e^{- 2 \chi z  } - 2 e^{- \chi z } \right ]
     &, z  < {\bar z} \\
    V_C(z)= - \frac{C_4}{(l + z) z^3}
     &, z \geq  {\bar z} \end{array}
  \right .
\label{vpot}
\end{eqnarray}
where $C_4 = C_3 l$,  $C_3$ being the vdW coefficient  and $l$ a characteristic length
which determines  the transition from the vdW ($z \ll l$) to the Casimir($ z \gg l$) regime.
The matching point ${\bar z}$ is determined by imposing the continuity condition for the interaction potential
($V_M({\bar z}) = V_C({\bar z})$) and its first derivative ($V_M'({\bar z}) = V_C'({\bar z})$).
%
%
%
%
The range of variation of the $C_3$ parameter is usually known so that
the only real free parameter of the Morse potential in this model is the stiffness parameter
$\chi$ ($D$ is  determined from the matching point ${\bar z}$).

The periodic function describing the grating  is written as
\begin{equation}
h(x) = \sum_{n=- \infty}^{+ \infty} \prod \left ( \frac{x - n d}{a}\right )
\end{equation}
where  $a$ is the width of the strips and $d$ the period  with  $a < d$. The $\prod (y)$-function is the
so-called unit impulse function: $0$ for $|y| > 1/2$, $1$ for $|y| < 1/2$ and $1/2$ for $|y|=1/2$.
In terms of a Fourier series, $h(x)$ is expressed as
\begin{equation}
h(x) =  \sum_{n=- \infty}^{+ \infty} c_n e^{i 2 \pi n x/d}
\label{hfun}
\end{equation}
with $c_0= a/d$, $c_{-n} = c_{n}$ and  $c_n=  (a/d) sinc (n a /d)$,
%
%
and $sinc (x) = sin(\pi x)/ \pi x$. When $d = 2 a$ (as in the experimental grating of Ref.
\cite{zhao1}), the terms beyond the sixth order are in practice no longer significant.
%
%
The periodic interaction potential can then be expressed as
\begin{equation}
U(x,z)  =   \sum_{n=- \infty}^{+ \infty} V_n (z) e^{i \frac{2 \pi n x}{d}}
\label{Vfourier}
\end{equation}
where the first term ($n=0$) is the interaction potential $V_0(z) = V(z)$
(see Eq. \ref{vpot}) and the coupling terms ($n \neq 0$) are given by
\begin{equation}
V_n (z) = 2  sinc (n \frac{a}{d}) V(z)  .
\label{coupling}
\end{equation}
%

As  has been recently shown \cite{salva2}, the elastic scattering of the incident
particles with the grating is theoretically well described by the CC formalism which accounts
for the quantum reflection probabilities as well as diffraction patterns.
%
%
%
%
The CC  equations are
\begin{equation}\label{close_coupling}
 \left[ \frac{\hbar^2}{2m} \frac{d^2}{dz^2} + \frac{\hbar^2}{2m} k_{n,z}^2
- V_{0}(z) \right] \psi_{n}(z) ~=~ \sum\limits_{n \neq n'}
V_{n - n'}(z) \psi_{n'}(z)
\end{equation}
with  $\frac{\hbar^2}{2m} k_{n,z}^2$  being the z-component of the kinetic energy of the
scattered particles.  The square z-component of the wave vector is written as a kinematic relation according to
\begin{equation}\label{tr}
 k_{n,z}^2 ~=~ k_i^2 - \left( k_i \sin \theta_i + \frac{2 \pi n}{d} \right)^2.
\end{equation}
with $\theta_i$ measured with respect to the normal to the surface.
Thus, when comparing with experimental results, theoretical positive $n$ diffraction orders
correspond to experimental negative  ones. For every $n$, the effective potential
$V_{0}(z)+ \frac{\hbar^2}{2m} (k_i \sin \theta_i + 2 \pi  n / d)^2$ in Eq.(\ref{close_coupling})
represents a diffraction channel, whose asymptotic energy is given by the second term.
This energy depends on $n$ and the incident scattering conditions (incident energy and polar
angle). Open (closed) diffraction channels have a positive (negative) normal kinetic energy
$\hbar^2 k_{n,z}^2/(2m) $. The coupling between channels $V_{n - n'}(z)$ is
given by Eq. (\ref{coupling}) since $n-n'$ is always an integer number.
The diffraction probabilities are obtained by solving the CC equations (Eq. \ref{close_coupling})
with the corresponding boundary conditions \cite{salva3}.

The diffraction intensities or probabilities, obtained by solving the CC equations given by Eq. (\ref{close_coupling}) 
with the usual boundary conditions \cite{salva2}, are expressed as
\begin{equation}
I_n = |S_{n0}|^2
\end{equation}
where $S_{nn'}$ are the elements of the scattering matrix , which give the amplitude of probability for an incident
wave at the specular channel ($n' =0$) and exiting by any of the open diffraction channels labelled by $n$. By construction, the $S$-matrix 
is unitary. It should be stressed, contrary to some
claims in the literature \cite{stickler}, that one should not ignore the closed channels in the computation. Although they are not
important in the asymptotic region where the coupling to them vanishes, they do affect strongly the diffraction probabilities as
well as the total reflection probability. This is but another indication that quantum reflection is
determined by the global potential and not only by the asymptotic form. This implies that numerical
convergence needs to be verified not only with respect to the grid size and integration step but also
to the number of closed channels.

In diffracting systems, when a diffraction channel becomes just open or closed, an emerging or evanescent
beam is observed, respectively. Due to the unitarity condition of the $S$-matrix, the diffraction intensities
undergo, in general, some abrupt variations. The corresponding  kinematic condition (\ref{tr}) fulfills $ k_{n,z}^2 = 0$. For
a given incident energy and $n$ value, the corresponding incident angle $\theta_i$ is called
the $n$-th order Rayleigh angle and this abrupt variation of the intensity is called emerging beam resonance or Rayleigh-Wood
anomaly \cite{prl} in grating scattering (or threshold resonance in atom-surface scattering \cite{tr1,tr2}).

The interaction potential given by Equation (\ref{vpot}) displays
classical turning points due to the repulsive part of the Morse potential. To distinguish between
quantum reflection and the "normal" reflection from the inner repulsive part
of the Morse potential, we impose absorbing boundary conditions
\cite{miller,muga} in the inner part.  For this purpose, a Woods-Saxon (WS) potential is
introduced as an imaginary part of the diffraction channel potentials
\begin{equation}
V_{WS} = \frac{A}{1 + e^{\alpha  \chi (z-z_i)}}
\label{17}
\end{equation}
which is essentially zero in the physically relevant interaction region and turns on sufficiently rapidly
but smoothly at the left edge of the numerical grid for the integration to absorb the flux.
The free parameters $A$ and $\alpha$ of the WS potential depend on the system under
consideration. Due to this numerical procedure, the resulting scattering matrix
${\bar S}$ is no longer unitary. The diffraction intensities are then given by ${\bar I}_n = |{\bar S}_{n0}|^2$
and the total quantum reflection probability is calculated from
\begin{equation}\label{qrp}
P^{QR} = \sum_n |{\bar S}_{n0}|^2   < 1
\end{equation}
for each initial condition. Due to the absorbing potential the theoretical diffraction efficiencies are defined as the ratio of the diffraction intensity 
${\bar I}_n$ to the total quantum reflection probability $P^{QR}$ rather than to the total incident flux.

\begin{figure}
\begin{center}
\includegraphics[scale=0.4,angle=0]{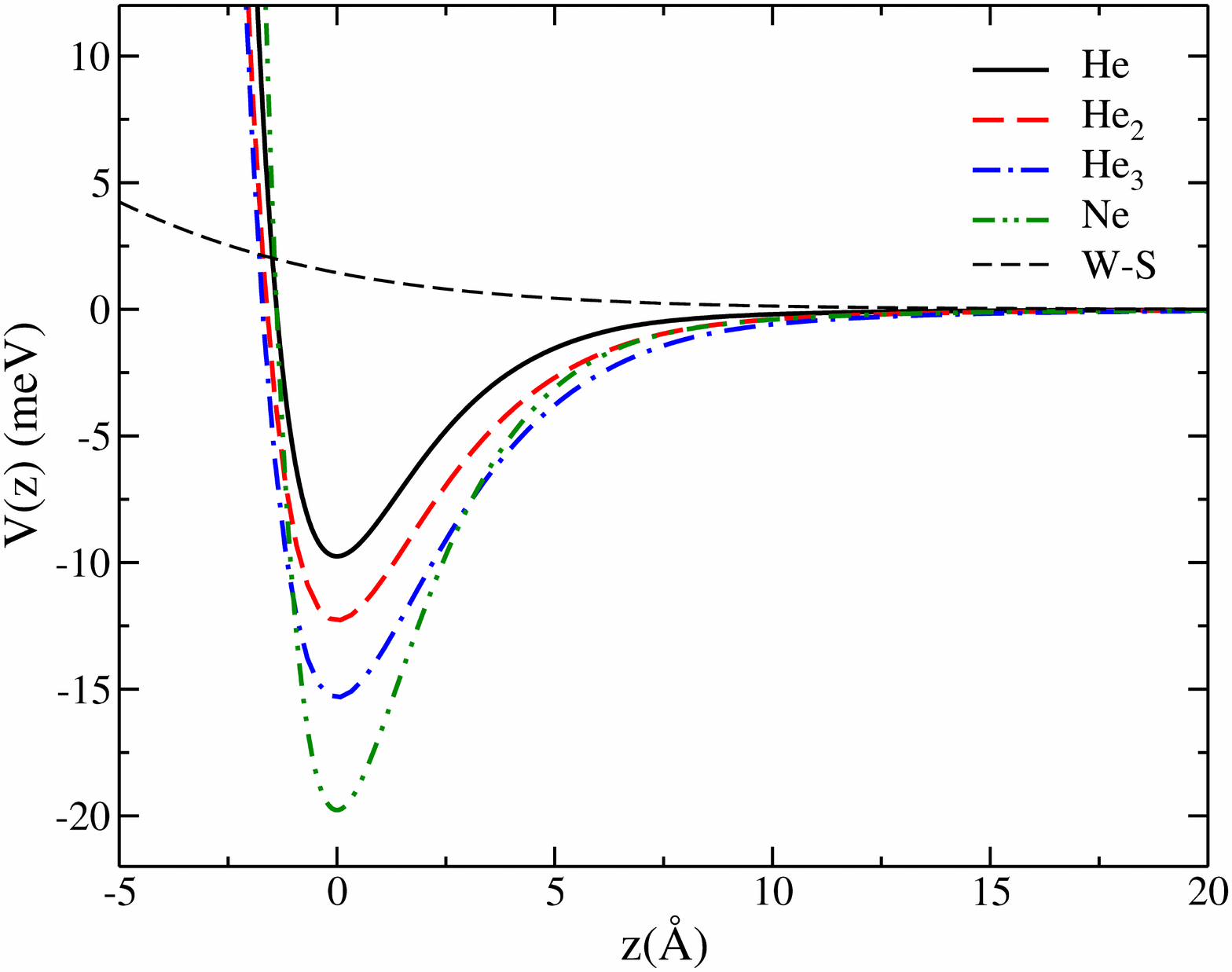}
\vspace{3.0cm}
\caption{(Color online) Vertical interaction potentials $V(z)$ between the incident particles and the grating.
The black dashed curve is the WS absorbing potential used. This potential is added as an imaginary part
for all diffraction channels with slightly modified parameters (see text).}
\label{pot}
\end{center}
\end{figure}
%

\section{ Results and discussion} 

As previously discussed, \cite{salva2} numerical convergence when solving the CC equations
is reached in two steps. First, the grid parameters have to be established. For the four incident particles under
study, the initial grid point is between -10 $\AA$ (for He)  and -20 $\AA$ (for He$_3$) and the
final integration point is between 500 $\AA$ (for He) and 1,000 $\AA$ (for He$_3$) and 2,000 $\AA$
(for $Ne$), the number of points of the grid ranging between 10,000 and 20,000. In the second step,
the maximum number of open and closed diffraction channels are chosen to be 61
(from $n= -30 \cdots +30$).
Only one free potential parameter must be fitted for each diffractive system to reproduce the experimental
reflection probabilities. The different potential parameters are listed in Table \ref{table1} for each incident particle.
\begin{table}
\caption{Parameters of the interaction potential $V(z)$ for the incident particles $He$, $He_2$, $He_3$
and $Ne$. The stiffness parameter of the Morse potential, $\chi$, is a free parameter fitted to reproduce
the corresponding experimental results and $D$ is the well depth. The characteristic lengths $l$ and parameters $C_3$
are based on values reported in previous works \cite{wieland1}.}
\vspace{1cm}
\centering
\begin{tabular}{|c||c||c||c||c|}
  \hline       Parameters & He & He$_2$ &  He$_3$ & Ne   \\
  \hline      $\chi$ ($\AA^{-1}$) &  0.5 & 0.43 & 0.405 & 0.5 \\
  \hline      D (meV) & 9.8 & 12.28 & 15.3 & 19.8 \\
  \hline      l ($\AA$ ) & 93 & 93 & 93 & 118.4  \\
  \hline       C$_3$ (10$^{-50}$ J m$^3$) & 3.5  & 7.0 & 10.5 & 7.0  \\ \hline
\end{tabular}
\label{table1}
\vspace{1cm}
\end{table}
%
%
%
\begin{figure}
	\includegraphics[scale=0.4,angle=0]{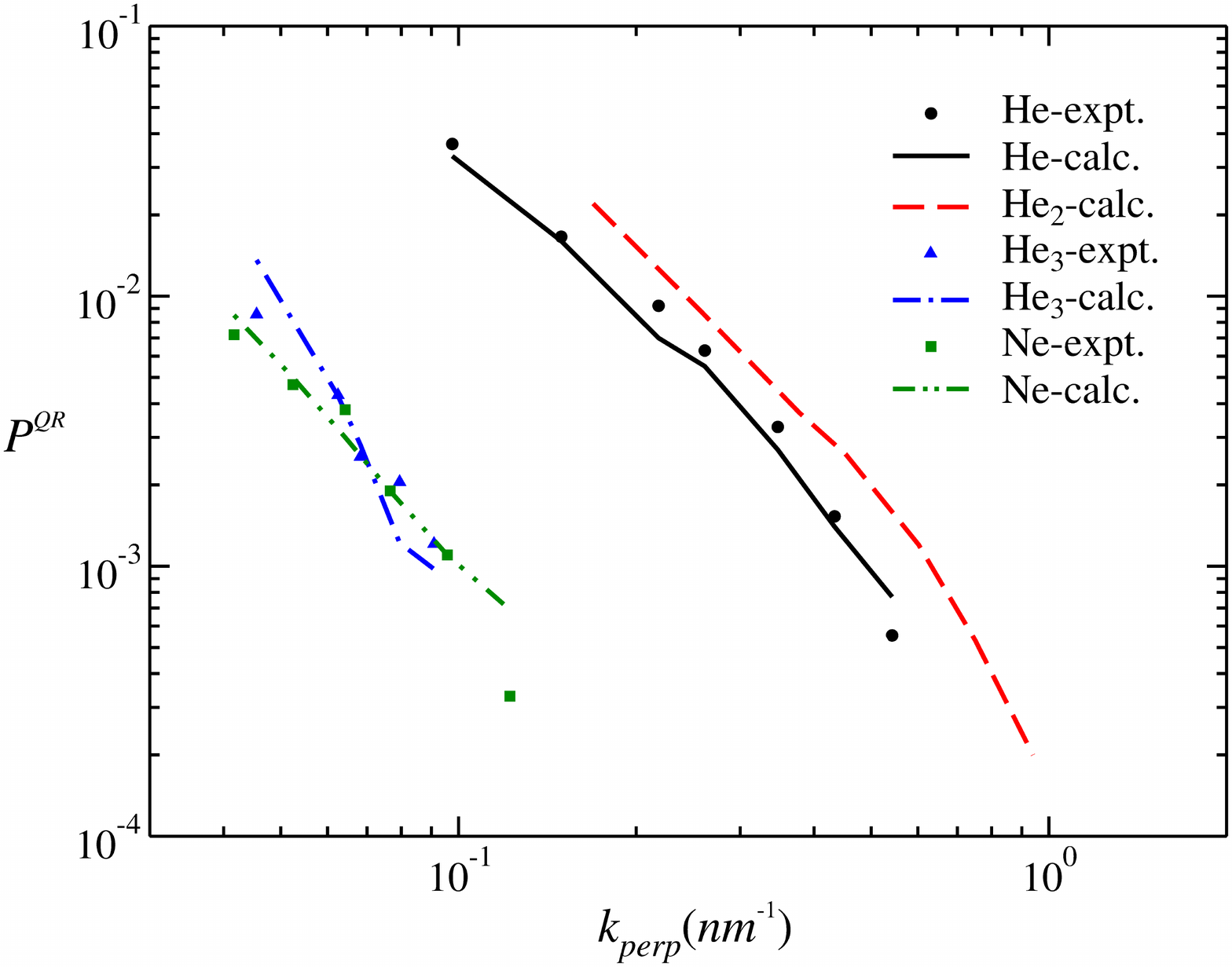}
	\caption{(Color online) Theoretical (multichannel calculations) quantum reflection probabilities for He
		($T_0 = 20$ K),  He$_2$  ($T_0 = 15$ K), He$_3$  ($T_0 = 8.7$ K) and Ne  ($T_0 = 40$ K)
		are plotted versus the perpendicular incident wave  vector $k_{perp}$ (in nm$^{-1}$). The
		results for the He dimer are not compared with experiment since they have been measured only 
		for scattering from a blazed grating rather than
       the "regular' grating used here. The experimental points have been provided by the authors of Refs. \cite{wieland1,wieland2}}
	\label{comparison}
\end{figure}
%
%
%
The $C_3$ and $l$ values are adapted from the expected range for He, He$_2$, He$_3$ and Ne  interacting with a transition metal surface
\cite{wieland1,zhao1}. According to Refs. \cite{zhao1,wieland1}, $C_3$ for He is $\sim 3.2-4.3 \, 10^{-50} \, J m^3$ 
since this is the expected range for the interaction with a transition metal surface. For He$_2$, one expects $l$ to be the same as He but $C_3$ 
to be two times larger. The same proportionality for $C_3$ is expected for He$_3$. This has been our guide to fit the interaction potentials for 
the three He clusters. For the Ne case, one expects larger values for $l$ and $C_3$ but our only guide has been the fitting to the corresponding
experimental results.The different potential parameters for the four systems used here seem to be  plausibly close to values obtained from
ab-initio arguments or calculations \cite{zhao1,wieland1}. The vertical interaction potentials $V(z)$ for the four systems and the 
WS potential are plotted in
Fig. \ref{pot}. As expected, the well depth of the Morse potential increases with the mass of the incident particle.

The WS potential (the dotted-dashed black curve in Fig. \ref{pot}) is "turned on"
in the region of the classical turning point of the interaction potential; in particular, at the initial
value of the grid $z_i$. Quantum reflection is not observed when the absorbing potential is placed
far to the right of the potential well,  where the vdW-Casimir tail of the interaction potential
is prevalent. In other words, the inner part of the interaction potential has a profound effect on
the reflection probability, showing that this region cannot be omitted from the $z$-grid integration.
This is again confirmed by the different values of the Morse potential parameters for the different incident particles.
All of the diffraction channels are modified by adding the imaginary  WS
potential. The couplings among them do not involve any imaginary part. The two parameters
$A$ and $\alpha$ needed to define the WS potential for every diffraction channel are chosen
after several runs of the  CC code.  In a  first run, those parameters are varied by including only
three diffraction channels: the specular channel ($n=0$) and  the two diffraction channels
labelled by $n = \pm 1$, with the demand being that the specular reflection probability is  similar to
that of the  one-channel calculation. In a second run, the next two diffraction channels $n=\pm 2$
are added by using the same criterion. This general numerical procedure is relatively straightforward to implement
because the corresponding parameters of the remaining channels are much less sensitive to the
total quantum reflection. The resulting parameters are listed in Table \ref{table2}.
\begin{table}
	\caption{The parameters $A$ and $\alpha$  of the WS potential used for the incident particles He, He$_2$, He$_3$
		and Ne. The initial grid points are $z_i = -10, -20, -21, -12$, repectively. The parameters for each system are written 
		as $(A_n,\alpha_n)$ where the subindices $0$ and $1$ are used for the specular channel ($n=0$) and  the first order diffraction 
		channels $n=\pm 1$ respectively and the subindex $2$ is used for the remaining diffraction channels. $A$ values are in atomic units.
	Numbers in parenthesis mean powers of ten; for example $7.0 (-4) \equiv 7.0 \times 10^{-4}$.}
	\vspace{1cm}
	\centering
	\begin{tabular}{|c||c||c||c||c|}
		\hline       Parameters & He & He$_2$ &  He$_3$ & Ne   \\
		\hline      $(A_0, \alpha_0)$ &  (7.0 (-4), 0.5) & (2.0 (-6), 0.1) & (2.0 (-3), 0.3) & (2.0 (-2), 0.9) \\
		\hline      $(A_1,\alpha_1)$ & (9.0 (-5), 0.1) & (9.0 (-5), 0.1)  & (2.0 (-1), 0.5) & (2.0 (-2), 1.5) \\
		\hline      $(A_2,\alpha_2)$ & (7.0 (-3), 0.3) & (4.0 (-2), 0.3) & (2.0 (-4), 0.1) & (2.0 (-2), 0.12)  \\ \hline
	\end{tabular}
	\label{table2}
	\vspace{1cm}
\end{table}
%

%
\begin{figure}
	\includegraphics[scale=0.4,angle=0]{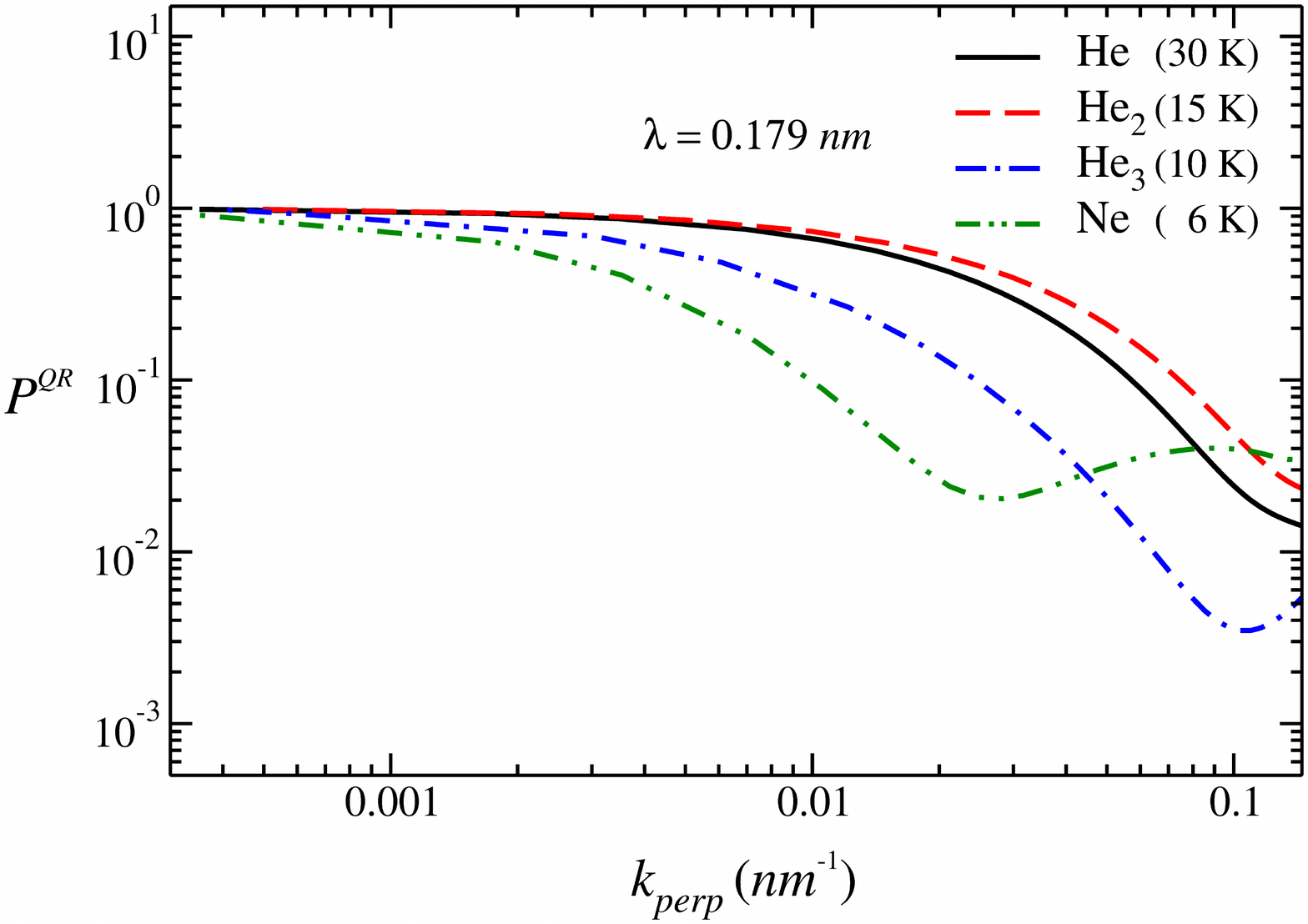}
	\caption{(Color online) Quantum reflection probabilities are plotted versus $k_{perp}$ (in $nm^{-1}$) for
		He, Ne, He$_2$ and He$_3$ at the same de Broglie wave length $0.179 \, nm$ . The source temperature
		$T_0$ used for each system is given in	parentheses. Note that the reflection probabilities only display
		a universal linear behavior at very small  $k_{perp}$. At higher values $k_{perp}$, the behavior is quite
		similar for the heavier masses (He$_3$,Ne) and lighter particles (He, He$_2$).}
	\label{universal}
\end{figure}
%

%
%

In Figure \ref{comparison}, the quantum reflection probabilities are plotted  versus
the perpendicular incident wave vector $k_{perp}$ (in nm$^{-1}$)  for: (i) He and He$_2$ with incident
energy $T_0= 20$ and $T_0 = 15$ K, respectively, with $\theta_i = 3.4, 5.2, 7.6, 9.1,12.1, 15.1, 18.9$ mrad,
(ii) He$_3$ with incident energy $T_0 = 8.7$  K  and $\theta_i = 0.8, 1.1, 1.2, 1.4, 1.6$ mrad and
(iii) Ne with incident energy $T_0 = 40$ K and $\theta_i = 0.5, 0.6, 0.7, 0.8, 1.1, 1.3$ mrad.
Black labels represent the experimental values of Ref. \cite{zhao1} and color curves correspond to
the multichannel calculation. The overall agreement between theory and experiment is fairly good.
The results corresponding to the He dimer are a prediction since no experimental information
exists in the literature for the regular grating used in our computations. The absorbing potential 
has negligible effects on the theoretical quantum reflection probabilities.

As clearly seen in this figure, the reflection probabilities at a given value of $k_{perp}$ are much smaller for
massive incident particles. They are quite similar for He and He$_2$ on the one hand  and He$_3$ and Ne,
on the other hand. Experimental confirmation would be highly desirable  to  validate the
theoretical behavior of the He dimer. As previously reported \cite{salva2}, the full interaction region (inner
and external regions) contributes coherently and equally to the reflection probabilities. In particular, at higher
perpendicular wave vectors incident particles  explore deeper and deeper regions of the potential well,
the attractive part  is no longer described by the vdW-Casimir tail. This part is considered
in our model by means of a Morse potential which is much more appropriate.

%
\begin{figure}
	\includegraphics[scale=0.4,angle=0]{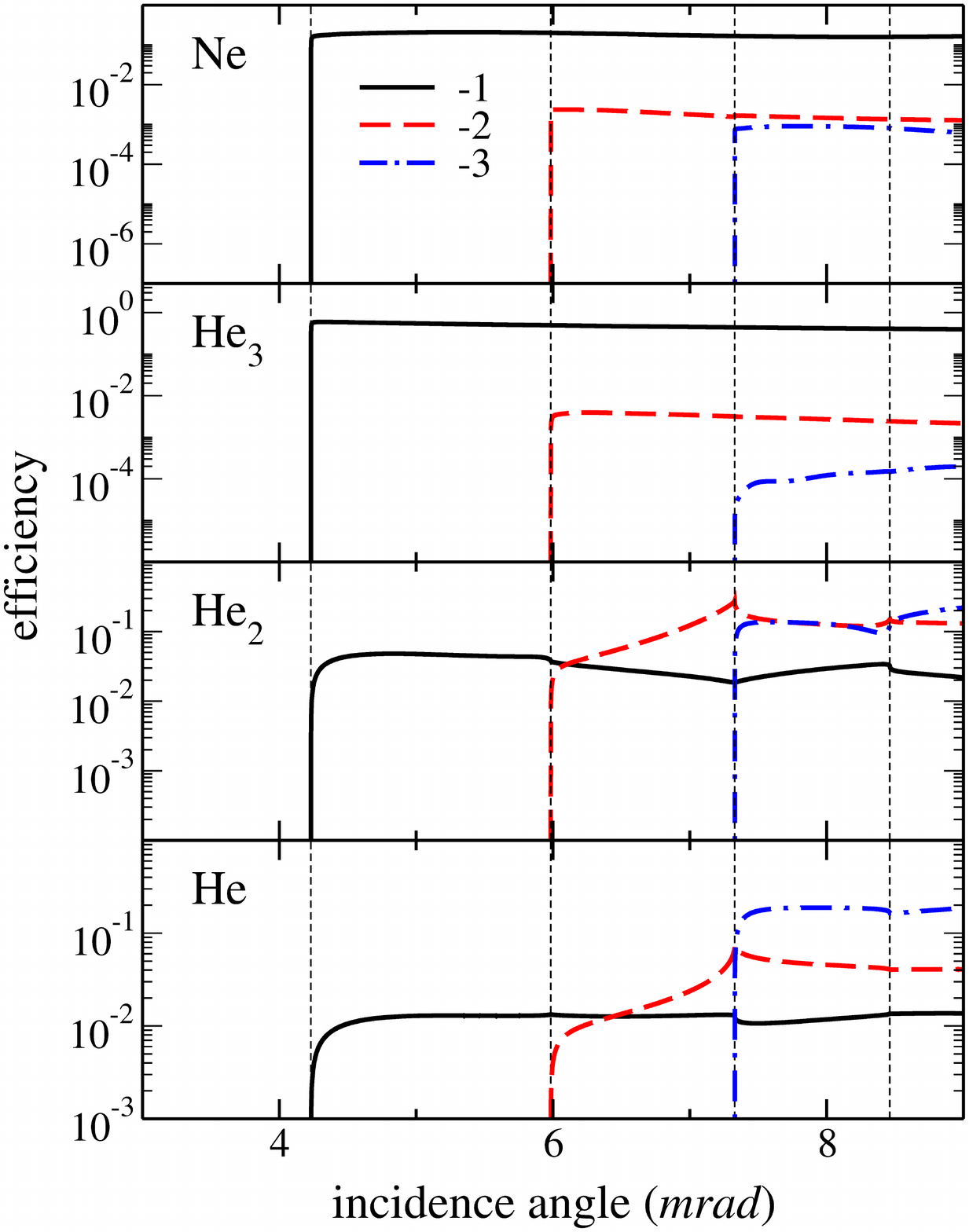}
	\caption{(Color online) Diffraction efficiencies are plotted versus the incident angle (in $mrad$) for
		He, Ne, He$_2$ and He$_3$ at the same de Broglie wavelength of $\lambda = 0.179 \, nm$.
		The Rayleigh angles are plotted as dashed vertical lines to indicate the opening or closing of a
		given diffraction channel.}
	\label{efficiencies}
\end{figure}

%
\begin{figure}
	\includegraphics[scale=0.4,angle=0]{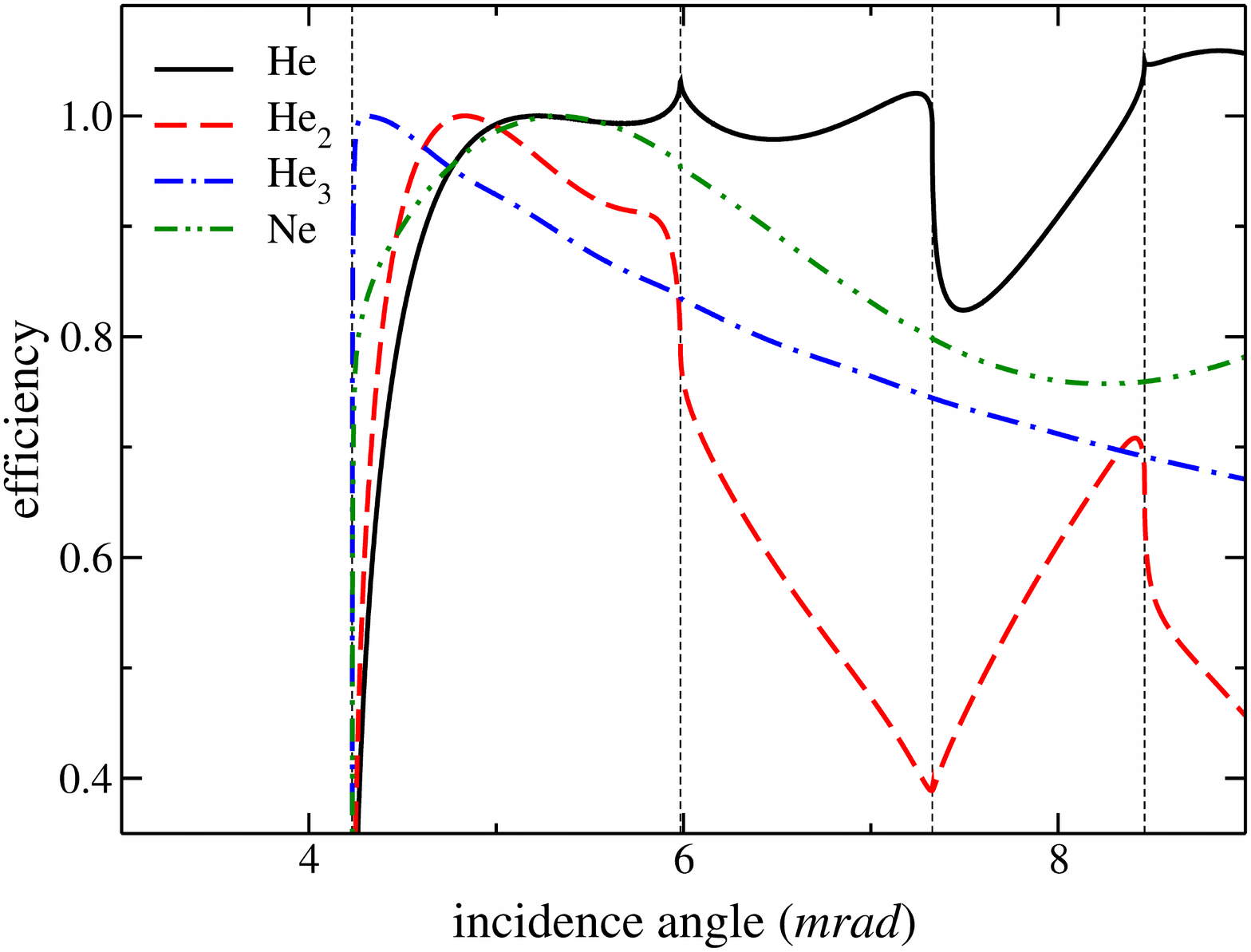}
	\caption{(Color online) Diffraction efficiencies of the open $n=-1$ diffraction channel are plotted versus the incident angle (in $mrad$) for
		He, Ne, He$_2$ and He$_3$ at the same de Broglie wavelength of $\lambda = 0.179 \, nm$.
		The Rayleigh angles are plotted as dashed vertical lines to indicate the opening or closing of a
		given diffraction channel.}
	\label{efficiencies-1}
\end{figure}

The  expected linear dependence of the quantum reflection probability on the incident perpendicular wavevector is revealed only at very
small incident perpendicular wave vectors as seen in Fig. \ref{universal}. The  theoretical (multichannel
calculations) quantum reflection probabilities are plotted versus $k_{perp}$ (in $nm^{-1}$) for He,
He$_2$, He$_3$ and Ne {\it but keeping the same  incident de Broglie wavelength fixed} at $\lambda = 0.179 \, nm$.
The source temperature $T_0$ used for each incident particle is given in parentheses. Here too, the 
slopes in the linear regime depend on the incident mass, the lighter the particle the smaller is the slope. 
As previously mentioned, the characteristic length $b$
governing the slope of the linear regime  is determined in a region where the potential differs appreciably from zero.
The linear behavior  is a direct result of boundary conditions and continuity of the wave function and its derivative and is thus "universal". However the magnitude of the slope is system specific and depends globally on the interaction potential. 
The linear regime is no longer observed when increasing the perpendicular wave vector showing a new functional dependence
with the wave vector. \cite{senn,jakob}

When the same incident wave vector is used for the four diffractive systems, it is also quite illustrative to demonstrate the universal dependence 
of the Rayleigh angles on the incident angle. \cite{wieland1} In Figure \ref{efficiencies}, diffraction efficiencies
are plotted as a function of the incident angle (in $mrad$) for the different incoming particles as well as the
Rayleigh angles (in dashed lines) showing when  the diffraction channels -1, -2, -3 just become open.  This occurs for any integer value of $n$ 
at which the kinematic relation (Eq. \ref{tr}) exactly vanishes. As may be inferred from the kinematic relation, the Rayleigh angles are then 
functions of the de Broglie wavelength of the incident particle and the lattice length only. Therefore, when scattering different particles on the 
same grating, one should expect that the threshold angles are only a function of the de Broglie wavelength. This is clearly seen in 
Fig. \ref{efficiencies}. At the Rayleigh angles, abrupt changes of the diffraction intensities are observed for the four systems (four panels)
due to the redistribution of the intensities among the open channels. This is also the typical behavior displayed in atom-surface
scattering. \cite{salva1}

It has also been observed in Ref. \cite{wieland1} that when plotting the efficiency of the $n=-1$ diffraction peak as a function of the incident 
angle on a logarithmic scale that it is only a function of the de Broglie wavelength of the incident particle. This was observed using a blazed grating. 
In Fig. \ref{efficiencies-1} we plot the efficiencies we obtained for the diffraction channel $n=-1$ on a linear scale as a function of the incidence 
angle, keeping the incident de Broglie wavelength fixed and normalizing the plots for the different particles to unity at their maximum. It is clear 
that our theoretical results do not show any "universal" behavior in this context. The dependence on the angle of incidence changes when 
using different particles and is determined by the overall potential and identify of the incident particle.

\section{Concluding remarks}

We have demonstrated, using the closed coupled formulation for scattering and employing absorbing boundary conditions that it is possible to 
account theoretically for the quantum reflection probabilities found experimentally when scattering He, He$_3$ and Ne from a regular grating. 
We have also used the same to predict the quantum reflection probabilities expected for the He$_2$ dimer. The same computations were then 
used to verify the universal dependence of the so called emerging resonances, or Rayleigh angles and their dependence on the incident de 
Broglie wavelength alone, as also observed experimentally in Ref. \cite{wieland1}. Finally, we have shown that one should 
not expect an universal dependence of the diffraction efficiency on the incident angle and wavenumber.

The physics underlying these results is we believe quite clear.  Quantum reflection is a coherent, nonlocal
phenomenon where all the regions (inner and outer) of the interaction potentials must be considered on equal
footing. In particular, the boundary condition which imposes the vanishing of the wavefunction
for sufficiently small distances from the grating implies that the quantum  reflection has very little to do with the so called badlands region of the long range attractive potential. The de Broglie wavelength of the incident particle is much longer
than the spatial extent of the badlands region. It is this large wavelength which is critical
for understanding the quantum reflection phenomenon. The wavefunction covers all regions of the potential so that
all of them affect the final reflection probability. The Morse potential used for describing the short-range interaction
is a convenient way to cover the whole interaction region. Quantum reflection does not occur at
tens or hundreds of nanometers away from the grating. Although the maximal density of the outgoing wavefunction is found very far from the surface, as in many other quantum phenomena, the wavefunction at any position is  affected by any other position. Just as in the two slit problem, the wavefunction far away from any of the slits is affected by the wavefunction at the slits, so here, the wavefunction at its maximum, which is far from the surface is affected by the wavefunction close to the surface.

The implication of this is that as observed in our computations, the magnitude of the linear slope of the reflection probability as a function 
of the incident perpendicular wavelength is system and potential specific. Similarly, the diffraction efficiencies are found to be system specific. 
Only the emerging resonances are universal since they are a kinematic effect, which signal the opening up of new diffraction channels. The
main difference between our computations and the experimental results presented in Ref. \cite{wieland1} is that we use a regular grating rather 
than a blazed grating. Only quantitative differences in the final results are expected when using a blazed surface.
In this context we note that the theoretical results presented here for the He$_2$ dimer are a prediction which could be validated using 
new experiments with a regular grating rather than a blazed one. There is no reason a priori to prevent using more 
sophisticated interaction potentials in this formalism as used in Ref. \cite{dalvit}.

\vspace{1cm}

{\bf Acknowledgment}: The authors would like to thank W. Sch\"ollkopf and B. S. Zhao for
providing us with their experimental results. This work is supported by the Programa Nacional de Ciencias Básicas de Cuba 
PNCB: P223LH001-108 (G.R.L. and J.R.S.), by a grant with Ref. FIS2017-83473-C2-1-P from the
Ministerio de Ciencia, Innovaci\'on y Universidad (Spain) (S.M.A.) and by grants of the Israeli Science Foundation, the Minerva Foundation,
Munich and the German Israel Foundation for Basic Research (E.P.).



\end{document}